\documentclass[aps,prl,10pt,twocolumn,longbibliography,superscriptaddress]{revtex4-1}

\usepackage{csquotes}
\usepackage{graphicx}% Include figure files
\usepackage{dcolumn}% Align table columns on decimal point
\usepackage{bm}% bold math
\usepackage{hyperref}% add hypertext capabilities
\usepackage{gensymb}
\hypersetup{
    colorlinks=true,
    urlcolor= blue,
    citecolor=blue,
linkcolor= blue}

\usepackage{amsmath}
\usepackage{amssymb}

\begin{document}

\title{\textbf{Proposal for a nonadiabatic geometric gate with an Andreev spin qubit}}

\author{Mostafa Tanhayi Ahari}
\affiliation{Materials Research Laboratory, The Grainger College of Engineering, University of Illinois, Urbana-Champaign, IL 
61801, USA}
\affiliation{Department of Physics and Astronomy $\&$ Bhaumik Institute for Theoretical Physics, University of California, Los Angeles, California 90095, USA}

\author{Yaroslav Tserkovnyak}
\affiliation{Department of Physics and Astronomy $\&$ Bhaumik Institute for Theoretical Physics, University of California, Los Angeles, California 90095, USA}

\begin{abstract}
     We study a hybrid structure of a ferromagnetic insulator and a superconductor connected by a weak link, which accommodates
     Andreev bound states whose spin degeneracy is lifted due to the exchange interaction with the ferromagnet. The resultant spin-resolved energy levels realize a two-state quantum system, provided that a single electron is trapped in the bound state, i.e., an Andreev spin qubit. The qubit state can be manipulated by controlling the magnetization dynamics of the ferromagnet, which mediates the coupling between external fields and the qubit. In particular, our hybrid structure provides a simple platform to manipulate and control the spin qubit using spintronic techniques. By employing a modified Hahn spin echo protocol for the magnetization dynamics, we show that our Andreev spin qubit can realize a nonadiabatic geometric gate.   
\end{abstract}
\maketitle

\section{Introduction}
\begin{figure}
  \includegraphics[scale=.078]{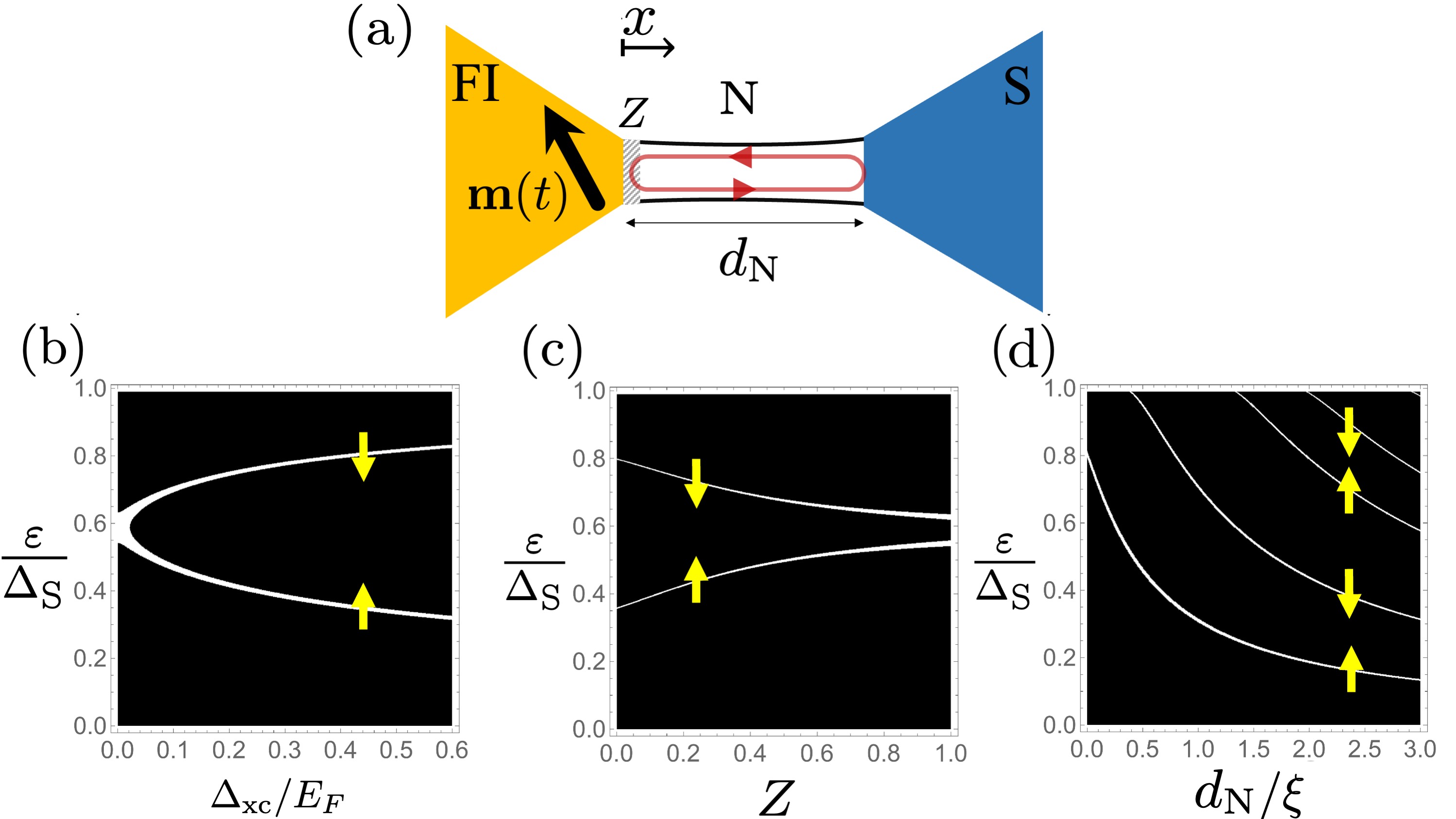}
   \caption{(a) A schematic for our hybrid structure realizing Andreev spin qubit due to the spin-resolved Andreev levels formed in the N region. (b) The spin-resolved ABS as a function of exchange field $\Delta_\text{xc}$ when $Z=0$ and $d_\text{N}/\xi=0.8$. (c) The spin-resolved ABS vs
   $Z$ when $\Delta_\text{xc}/\mu=0.6$ and $d_\text{N}/\xi=0.8$. (d) The spin-resolved ABS vs $d_\text{N}$, with $Z=0$ and 
   $\Delta_\text{xc}/\mu=0.6$. In these plots and throughout this paper, 
   we have chosen $V_0/\mu=1.7$ and $\Delta_\text{S}/\mu=0.003$.} 
   \label{AQ}
\end{figure}

Motivated by the prospect of scalable device 
fabrication and circuit design, the search for two-state quantum systems or qubits in solid-state systems has been a major 
experimental and theoretical endeavor~\cite{Petta,task1,*task2,*task3,*task4,*task5}. Among all,
Andreev physics-based qubits present a particularly promising route.    
In the superconducting state at low temperatures, the Fermi-level degrees of freedom are frozen out. 
Therefore, most of the dissipative mechanisms are eliminated so that the qubit  
may exhibit long coherence times~\cite{Ketterson1,*Ketterson2,Devoret,Urbina}. 
In this paper, we study a hybrid structure of a ferromagnetic insulator (FI)
connected by a weak link (a normal region N) to an $s$-wave superconductor (S) that accommodates spin-resolved 
Andreev bound states (ABS) in the weak link [see Fig.~\ref{AQ}(a)]. We show that a pair of spin-resolved 
ABS, when occupied by an electron, creates an 
Andreev spin qubit (ASQ). On account of a large level spacing in our ASQ, which is due to the exchange field of the FI, 
the external fields controlling the FI magnetization are strongly coupled to the qubit spin. Due to the strong 
coupling, we expect that 
spin-flipping errors originating from weak random fields in the environment, such as in the case of hyperfine coupling to 
nuclei~\cite{Loss21,*Loss22,*Loss23}, will be suppressed.   

A hallmark feature of the ABS formed in a Josephson junction is that the occupied levels
modify a supercurrent across the junction~\cite{Nazar,Manfra} so that transitions between Andreev levels can be detected 
by means of transport measurement. This is the physical basis of Andreev level qubits~\cite{Urbina,Zaz1,*Zaz2,*Zaz3}. It was also 
observed that in spin active weak links, e.g., with spin-orbit coupling~\cite{Nazar,Devoret}, or magnetic impurity~\cite{ASQ1,*ASQ2},
the spin degeneracy of the levels can be lifted, so that the supercurrent flow would depend on  
the spin state of the electron trapped in the bound state. In our ASQ, the exchange field of the FI breaks the spin degeneracy of the levels, so that
quantum information can be stored in 
the spin state of the electron~\cite{Shum}. Similarly, to detect or read out the qubit state, we study the superconducting current affected by the occupied 
spin-resolved levels.

Generically, 
the unwanted qubit coupling to the environment and inaccurate external control can reduce the qubit state coherence time, which hinders the experimental implementation of scalable qubit designs~\cite{Petta,task1,*task2,*task3,*task4}. 
The qubits with gates based on the geometric phases may have the inherently fault-tolerant advantage due to the fact that the geometric 
gates depend only on some  global  geometric features of the evolution but 
independent of evolution details, so that they can be robust against control errors~\cite{PRLM1,*PRLM2,GG1,*GG2,*GG3,*GG4,Tong}. 
Moreover, utilizing nonadiabatic gates with high operation speed may reduce exposure time to
the environment, rendering a high-fidelity gate~\cite{Xue}. Here, we will show that our hybrid structure is a natural platform to realize  
a nonadiabatic geometric gate.

\section{ The hybrid structure with spin-resolved ABS}
Consider the FI-N-S hybrid structure with the following Bogoliubov–de Gennes Hamiltonian
\begin{align}\label{BdG}
   \hat{H}_\text{FINS}=\begin{pmatrix}
   H_0(p,{\bf m}) &  i\Delta_\text{S}(x) \, \sigma_y \\
    -i \Delta_\text{S}(x) \, \sigma_y & -H^*_0(-p,{\bf m})
   \end{pmatrix},
\end{align}
where $H_0(p,{\bf m})=\frac{p^2}{2m}-\mu+V(x)+\Delta_{\text{xc}}(x){\bf m}(t)\cdot \boldsymbol{\sigma}$, $\mu$ is the chemical potential,  $\Delta_{\text{xc}}(x)=-\Delta_{\text{xc}}\, \Theta(-x)$ is the magnetic
exchange field, and $\Delta_\text{S}(x)=\Delta_\text{S}\, \Theta(x-d_\text{N})$ is the superconducting pair potential,
both written in terms of the Heaviside step function. 
The scalar potential $V(x)=V_0 \,\Theta(-x)+\hbar v_F Z\delta (x)$ models 
the FI electron gap where $V_0> (\mu+\Delta_\text{S}+\Delta_{\text{xc}})$, and $Z$ parametrizes an interface barrier 
potential where $v_F$ is the Fermi velocity. The unit vector ${\bf m}(t)$ gives the time-dependent direction of the magnetization, 
and $\boldsymbol{\sigma}=(\sigma_x,\sigma_y,\sigma_z)$ 
are Pauli matrices operating in spin space. 

Our goal is to find the subgap ($|\varepsilon|<\Delta_\text{S}$) energy spectrum of bound states, which are confined in the normal region. Imposing a condition for constructive quantum interference in a McMillan-Rowell process for the electrons\textemdash 
four times crossing N with two Andreev conversions as well as two reflections from FI, once as electron and once as hole~\cite{Eschrig}\textemdash results in a transcendental equation 
\begin{align}\label{RW}
   \varphi_\sigma(\varepsilon) +2\theta_{\text{A}}+4\theta_{\text{N}}=2\pi n, \,\,\, n=0, \pm1,\cdots ,
\end{align}
whose solutions are the discrete bound states energy levels $\varepsilon$ (measured from $\mu$). 
Here, $\sigma\in\{\uparrow,\downarrow\}$ is specified with respect to the FI magnetization direction, 
$\varphi_\sigma(\varepsilon) \equiv \vartheta_\sigma({\varepsilon})-\vartheta_{-\sigma}(-\varepsilon)$, with 
$\vartheta_\sigma({\varepsilon})$ being the spin-dependent phase acquired by the electron upon reflection from the FIN interface, $\vartheta_\sigma(\epsilon)= \vartheta_\sigma^{(0)}(\varepsilon)+2\,\text{arg}\big(1-iZ-iZe^{-i\vartheta_\sigma^{(0)}(\varepsilon)}\big)$, where $\vartheta^{(0)}_\sigma(\varepsilon)\equiv\vartheta_\sigma({\varepsilon,Z=0})$~\cite{Eschrig}. $\theta_{\text{N}}=d_\text{N}\varepsilon/\hbar v_F$ where $d_\text{N}$ is the normal region thickness. 
Since for subgap energies the S supports only Cooper pair tunneling into the S,
the incident electrons from the N region on the clean N-S interface 
reflect back as a hole~\cite{ratio} (experiencing an Andreev reflection), where the phase change associated with the Andreev reflection is given by 
$\theta_{\text{A}}=- \arccos{(\varepsilon/\Delta_\text{S}})$. 

When $d_\text{N}$ is smaller than the superconducting coherence length $\xi$ and $\Delta_\text{xc}=0$, Eq.~\eqref{RW} can admit 
a single spin-degenerate positive-energy solution whose degeneracy is 
removed by increasing $\Delta_\text{xc}$ [see Fig.~\ref{AQ}(b)]. We point out that an oxide layer formed at the FIN
interface or a tunable 
mismatch between the electronic properties across the interface 
create an effective ultrathin insulating layer that can greatly affect 
the probability of electron and hole evanescent penetration into the FI region. 
To capture the essential effect of this insulating layer, we use an interface barrier potential with strength $Z$, which 
can be utilized to tune the effective exchange interaction experienced
by electrons and holes~[see Fig.~\ref{AQ}(c)]. Moreover, the number of ABS found in a clean N region depends on its thickness, 
that is, by increasing $d_\text{N}$ the ABS levels are pushed towards the 
middle of the gap at $\mu$, so that more energy levels begin to appear [see Fig.~\ref{AQ}(d)]. 

\section{The Andreev spin qubit}
Let us consider a hybrid structure with only two positive-energy 
spin-resolved ABS $(\varepsilon_1,\varepsilon_2)$, and ignore the continuum of positive energy levels above the S gap. We 
define the ground state of the system when these ABS are unoccupied and 
measure energies with respect to 
$(\varepsilon_1+\varepsilon_2)/2$. The energies then are $-(\varepsilon_1+\varepsilon_2)/2$ for the ground,
$(\varepsilon_1-\varepsilon_2)/2$ and $(\varepsilon_2-\varepsilon_1)/2$ for two spin-$\frac{1}{2}$, 
and $(\varepsilon_1+\varepsilon_2)/2$ for excited state. 
The ground and excited states have even (electron number) 
parity. The odd parity states, on the other hand, correspond to 
a single electron in the system realizing an ASQ. ABS wave functions with energies $(\varepsilon_1, \varepsilon_2)$ may have different spatial components. 
However, due to the tunneling treatment of the magnetic insulator, we assume that the ABS wave functions differ mainly in their spin character, with the orbital components being essentially identical. The effective ASQ Hamiltonian written in terms of the instantaneous quantization axis ${\bf m}(t)$ reads 
\begin{align}\label{AqH}
 H_\text{A}(t)=-\frac{1}{2}\Delta\, {\bf m}(t)\cdot\boldsymbol{\sigma}, 
\end{align}
where $\Delta\equiv \varepsilon_2-\varepsilon_1>0$. 
It is clear that the timescale for the coherent manipulation of the qubit, $t$, is set by the lifetime of the odd-parity sector, $\tau_e$. 
In the temperature regime considered in this paper, $ k_\text{B}T\ll \Delta_\text{S}$, the odd-parity sector
can be prepared, for example, by microwave irradiation~\cite{irra,Nazar}, electron injection by means of
voltage gates~\cite{Wendin}, or 
spontaneously when an electron is stochastically trapped in the bound state. The latter
mechanism, employed in the recent experiment~\cite{Devoret}, is due to ubiquitous
nonequilibrium quasiparticles and likely originated from a background 
stray photons with energy exceeding the threshold for breaking a Cooper pair~\cite{Houzet}. There are more deterministic approaches to preparing the ASQ in odd parity state~\cite{Dev2,Wes1,*Wes2}. For example, if the junction is irradiated with the microwave of frequency $\omega=(\varepsilon_1+\varepsilon_2)/\hbar$, the microwave drive can break a Cooper pair placing electrons on the Andreev levels $(\varepsilon_1,\varepsilon_2)$. Moreover, imposing the condition $\Delta_\text{S}-\varepsilon_2<\hbar \omega< \Delta_\text{S}-\varepsilon_1$, would guarantee that the microwave drive can only excite the electron from Andreev level $\varepsilon_2$ into the continuum, which results in the odd parity state~\cite{Dev2}.  

Once an electron is trapped, the probability of thermally activated parity-switching processes 
contains an exponentially small factor $\text{exp}(-\Delta_\text{S}/k_\text{B}T)$ for tunneling  
leading to the long lifetime of the trapped electron 
in the ABS, which is observed to exceed $ \tau_e=100$ $\mu$s~\cite{long-lived}. 
In general, when there are available subgap states caused by, e.g., spatial variations of the 
order parameter or impurities~\cite{Shiba}, $\tau_e$ will be finite provided the trapped electron 
can tunnel through the gap. We note that the microscopic details of the hybrid structure 
can greatly modify the residence time for the trapped electron, which can be determined from the \enquote{lifetime matrix}~\cite{lft}. 
As a result, the timescale for the coherent manipulation of the qubit reads
\begin{align}
    \hbar/\Delta_\text{S}< t < \tau_e,
\end{align}
where the lower bound is imposed to avoid dynamical mixing with the continuum states when the qubit state 
is evolving. 

\section{The qubit manipulation.}
Isolated single spins can be coherently 
manipulated using both electrical and optical techniques~\cite{Petta,spinc1,*spinc2}. In our hybrid structure, however, we accomplish the qubit manipulation by 
the coherent control of the magnetization dynamics. We study the qubit dynamics when the magnetization precession
is both resonant and nonresonant. 
When resonant, our hybrid structure enhances the Rabi oscillation frequency of the qubit dynamics.   
When nonresonant, as we discuss, our hybrid structure has the benefit of implementing
nonadiabatic single-qubit gates via spintronic techniques. Because of the experimentally achievable high-speed control of the magnetization dynamics~\cite{high-speed1,*high-speed2}, the nonadiabatic spintronic manipulation of the qubit state renders a shorter qubit evolution time, which is an important advantage in realizing high-fidelity quantum gates~\cite{Sai}.  

Consider first an FI dynamics at resonance, where the magnetization 
parametrized as ${\bf m}(t)=(\sin\theta\cos(\omega t+\phi_0) ,\sin\theta\sin(\omega t+\phi_0) ,\cos\theta)$ exhibits
precessional motion around the $z$ axis with
ferromagnetic resonance (FMR) frequency $\omega=\gamma h_0\equiv \omega_{\text{FMR}}$, where $\gamma$ is the gyromagnetic ratio, and $h_0$ is an effective 
the magnetic field in the $z$ 
direction. Note that $h_0$ may contain a static external field, 
demagnetization field, and other crystalline anisotropy fields~\cite{cone-dyn1,*cone-dyn2},
where the effect of the external magnetic field on the qubit, if nonzero, can easily be 
incorporated into Eq.~\eqref{RW} as a spin-dependent phase shift in the normal part.  
The cone angle for the magnetization is determined by a transverse rf (microwave) field,  $h_{\text{rf}}$, 
as $\theta\sim h_{\text{rf}}/\alpha h_0$, where $\alpha$ is 
the dimensionless Gilbert damping constant that parametrizes the inherent 
spin angular momentum losses of the magnetization dynamics.

The qubit spin, on the other hand, exhibits Rabi oscillations with frequency $\Omega\equiv\Delta \sin\theta/\hbar$ when 
$\omega=\Delta\cos\theta/\hbar$, which is the electron-spin-resonance (ESR) frequency. Now, matching the ESR and FMR frequencies and assuming 
$\theta\ll 1$, we get
\begin{align}
    \Omega\approx\frac{\gamma h_{\text{rf}}}{\alpha}.
\end{align}
For small damping $\alpha\ll1$, $\Omega\gg \gamma h_{\text{rf}}$. 
This shows that, for a given microwave stimulus $h_{\text{rf}}$, the magnetization acts as a mediator with the bonus of spatially
focusing and intensifying 
the external field that enhances the qubit Rabi oscillation frequency. 

\begin{figure}
  \includegraphics[scale=.084]{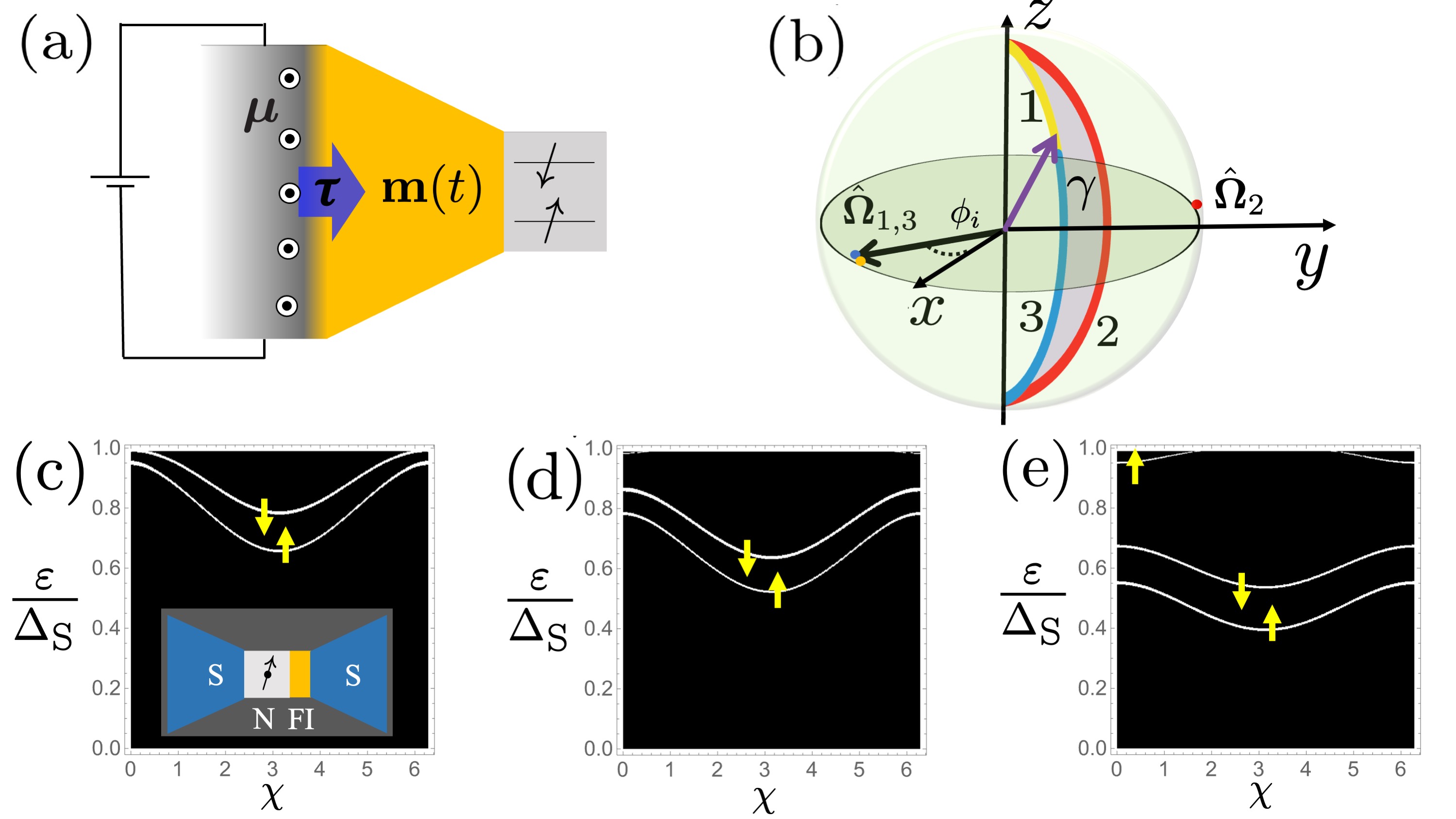}
   \caption{(a) Schematic of the FI in contact with two normal metals on the left and right. The spin-resolved ABS are localized in the right normal metal. The magnetization dynamics are coherently manipulated by torque $\boldsymbol{\tau}$ engendered from the out-of-the-plane spin accumulation $\boldsymbol{\mu}$ in the left normal metal. The spin accumulation $\boldsymbol{\mu}$, on the other hand, can be controlled by spin Hall effect~\cite{SA}.  
   (b) Bloch-sphere representation of $\langle g_+(t)|\boldsymbol{\sigma}|g_+(t)\rangle$, purple-color vector, undergoing a closed path.  
   The spin evolves through paths 1 (yellow), 2 (red), and 3 (blue) rotating about orientations $\boldsymbol{\Omega}_1,\boldsymbol{\Omega}_2$, and $\boldsymbol{\Omega}_3$, respectively. The solid angle associated with the enclosed (shaded) area is twice the accumulated Berry phase $\gamma$. The inset in (c) shows a schematic of the S-N-FI-S setup, where the supercurrent through the junction can be used to probe the spin state of the Andreev levels. (c, d, e) Spin-split ABS as a function of phase difference $\chi$ for $d_\text{N}/\xi=0.1, \,  0.4,$ and $0.8$, respectively. Here, we have set $Z_\uparrow=1$ and $Z_\downarrow=1.5$. } 
   \label{gg1}
\end{figure}

In contrast, when the FI dynamics are nonresonant, $\omega\ll\omega_{\text{FMR}}$, one can control phase $\phi_0$ to 
implement arbitrary single-qubit operations. In this regime, a natural spintronic way to maintain a magnetization precession at $\omega=\omega_{\text{ESR}}$ 
in the \enquote{conical} state is simply by maintaining an out-of-the-plane spin accumulation $\boldsymbol{\mu}$ in an attached normal metal [see Fig.~\ref{gg1}(a)]. The normal metal with spin accumulation can exert torque
$\boldsymbol{\tau}\sim {\bf m}\times (\boldsymbol{\mu}\times {\bf m}/\hbar-\partial_t{\bf m})$~\cite{Hector} 
on the magnet at the interface. Note that when $\boldsymbol{\mu}=0$ the resultant torque is the ordinary Gilbert 
damping endowed by the normal metal~\cite{TserPRL}. 
The presence of the spin accumulation can balance the damping torque leading to coherent precession of magnetization
at $\omega=\omega_{\text{ESR}}$ with 
$\boldsymbol{\tau}=0$. 
In this case, the magnitude of spin accumulation controls the frequency $\omega=|\boldsymbol{\mu}|/\hbar$, like in a
Josephson relation. 
As a result, phase changes could be accomplished simply by short pulses of large spin accumulations, which can be within the reach of current spintronic experimental techniques~\cite{high-speed1,*high-speed2}. 
To realize the conical state, one could take simply an easy-plane magnet. In a uniaxial crystal, the anisotropy energy 
contributes to the magnetic free energy with a term $\sim m_z^2$, which is minimized when 
the spontaneous magnetization direction lies in the $xy$ plane~\cite{Landau}. As a result, the out-of-the-plane angle can be controlled by 
a normal magnetic field. 

\section{ Nonadiabatic geometric gate}
As an illustrative application, 
one can adopt a protocol through which the qubit state accumulates only a geometric phase, i.e., the dynamic phase 
is zero during the whole evolution. To that end, we follow the protocol presented in Ref.~\cite{Tong} and show that
our ASQ serves as a natural platform to implement a nonadiabatic geometric gate.
In the rotating frame, when $\omega=\omega_{\text{ESR}}$, the ASQ Hamiltonian can be written as $H^{(r)}=-\frac{ \hbar}{2} \boldsymbol{\Omega}\cdot \boldsymbol{\sigma}$, where 
$\boldsymbol{\Omega}=\Omega\,(\cos\phi_0,\,\sin\phi_0,\,0 )$.  
Consider now $\boldsymbol{\Omega}$ to undergo a cycle 
described by the following three fixed orientations of $\boldsymbol{\Omega}$ connected by fast quenches:
\begin{eqnarray}\label{steps}
    \phi_1&&=\phi_0-\frac{\pi}{2} \hspace*{1.6cm} t_0< t \leq t_1,\nonumber\\
    \phi_2&&=\phi_0+\gamma+\frac{\pi}{2} \hspace*{1.0cm} t_1 < t \leq t_2,\nonumber\\
    \phi_3&&=\phi_0-\frac{\pi}{2} \hspace*{1.6cm} t_2 < t \leq t_3,
\end{eqnarray}
where $t_0\equiv0$. As a result, the full qubit evolution operator (or the quantum gate) reads
\begin{align}\label{gg}
U_g(t_3)&=e^{-\frac{i}{\hbar}\int_{t_2}^{t_3}H^{(r)}_{3}\,dt}e^{-\frac{i}{\hbar}\int_{t_1}^{t_2}H^{(r)}_{2}\,dt}e^{-\frac{i}{\hbar}\int_{0}^{t_1}H^{(r)}_{1}\,dt },
\end{align}
where $H^{(r)}_i=-\frac{\hbar}{2} \boldsymbol{\Omega}_i\cdot \boldsymbol{\sigma}$ with $ \boldsymbol{\Omega}_i=\Omega\,(\cos\phi_i,\,\sin\phi_i,\,0 )$. 
Now, if we impose $\Omega t_1\equiv \beta$, $\Omega(t_2-t_1)=\pi$, and $\Omega(t_3-t_2)=\pi-\beta$, the two orthogonal states $|g_\pm\rangle$, 
\begin{align}
     {\bf n}\cdot\boldsymbol{\sigma}|g_\pm\rangle &= \pm |g_\pm\rangle, \\
     {\bf n}&=(-\sin\beta \cos\phi_0,-\sin\beta \sin\phi_0,\cos\beta),\nonumber
\end{align}
undergo a particular cyclic evolution where the initial and final states are related by a purely geometric phase factor. To see this, one
can check that $\langle g_{\pm}(t)| H_i^{(r)} |g_{\pm}(t)\rangle=0$ for $i=1,2,$ and $3$, where $|g_{\pm}(t)\rangle=U_g(t)|g_{\pm}\rangle$. This implies that 
the dynamic phase at each stage of the evolution is zero. As a result, the geometric gate is obtained as
\begin{align}\label{gg}
U_g(t_3)&= e^{i\gamma {\bf n}\cdot\boldsymbol{\sigma}}=e^{i\gamma}|g_+\rangle \langle g_+| +e^{-i\gamma}|g_-\rangle \langle g_-|. 
\end{align}

 In order to elucidate the geometric nature of the angle $\gamma$,
we note that the expectation values of spin $\langle g_+(t)|\boldsymbol{\sigma}|g_+(t)\rangle$ at times $0, t_1, t_2$, and $t_3$ are ${\bf n},\hat{z},-\hat{z}$, and ${\bf n}$, respectively, undergoing a 
cyclic evolution on the Bloch sphere where the subtended solid angle associated with the enclosed path
is given by $2\gamma$ [see Fig.~\ref{gg1}(b)]. Notice that the case of $\gamma=0$ and $\beta=\pi/2$
is equivalent to Hahn spin echo~\cite{Hahn1,*Hahn2} with a $\pi$-pulse (in the second step). Thus, 
the protocol given in Eq.~\eqref{steps} can be considered as a generalized spin echo 
with a $(\pi+\gamma)$-pulse. 
Since the geometric phase depends on the evolution paths, quantum gates based on the geometric phases are 
resilient against errors in the evolution details, i.e., control errors
~\cite{Tong,Jones1,*Jones2,Zhang2}. 

\section{ The qubit readout}
To detect or read out the qubit state one can probe a small bias supercurrent, 
which requires the usage of another weakly coupled superconducting lead to the FI [see the inset in Fig.~\ref{gg1}(c)]. Josephson junctions consisting of an 
FI~\cite{spinf} are believed to have interesting properties, such as Josephson $\pi$ state~\cite{pi1,*pi2}. Here, to study the ABS in an S-N-FI-S Josephson junction, we model the FI layer as a thin insulating barrier with spin-dependent parameter $Z_\sigma$~\cite{Zs}. As a result, right- and left-moving electrons (and holes) get a spin-dependent coupling leading to spin-polarized ABS $\varepsilon_\sigma$. The discrete spectrum of these bound states can be related to the scattering matrix of the normal region~\cite{Beenaker} as
\begin{align}\label{scat}
\text{Det}\Big [ I-e^{i2\theta_{\text{A}}}r_A^*s_{e,\sigma}r_A s_{h,-\sigma} \Big]=0,
\end{align}
where $I$ is a two by two identity matrix, $r_A=\text{diag}(e^{-i\chi/2},e^{i\chi/2})$, the scattering matrix for electrons (holes) with spin $\sigma$, $s_{e(h),\sigma}$, is given by
\begin{align}
s_{e,\sigma}=t_{\text{N}}\begin{pmatrix}
    r_\sigma t_{\text{N}}^{-1} & t_\sigma \\
    t_\sigma & r_\sigma t_{\text{N}}
\end{pmatrix}
\end{align}
with $t_\text{N}= e^{i(\theta_\text{N}+k_F d_N)}$, $r_\sigma=-iZ_\sigma/(1+iZ_\sigma)$, $t_\sigma=1/(1+i Z_\sigma)$, and $s_{h,\sigma}(\theta_\text{N})=s^*_{e,\sigma}(-\theta_\text{N})$.
One can check that these Andreev levels satisfy 
\begin{align}\label{ABSJ}
     \cos(2\theta_{\text{A}}+2\theta_\text{N}+\theta) =\lambda(\chi,\theta_N),
\end{align}
where $\lambda(\chi,\theta_N)\equiv \sqrt{R_\uparrow R_\downarrow} \cos{(2\theta_\text{N})}+\text{sign}(Z_\uparrow Z_\downarrow)\sqrt{T_\uparrow T_\downarrow} \cos{\chi}$ with 
$R_\sigma=r^*_\sigma r_\sigma$ and $T_\sigma=t^*_\sigma t_\sigma$ are spin-dependent reflection and transmission probabilities due to the FI layer, $\theta=\theta_\downarrow-\theta_\uparrow$ with $\theta_\sigma=\arctan(1/Z_\sigma)$ being the phase shift of the reflected electron from the N-FI interface, 
and $\chi$ is the phase difference between the superconducting leads. Note that for $Z_\sigma\gg1$ Eq.~(\ref{scat}) produces a similar constraint to that given in Eq.~(\ref{RW}). Figures~\ref{gg1}(c)-\ref{gg1}(e) show the spectrum of ABS for different normal layer thicknesses. 
As a consequence of the broken spin degeneracy, the supercurrent can depend on the spin state of the occupied Andreev levels~\cite{Nazar}. 
For simplicity, we consider the limit of $d_\text{N}\rightarrow0$, where spin-dependent supercurrent $I_\sigma\sim d \varepsilon_\sigma /d\chi$ can be written as 
\begin{align}
    I_\sigma\sim \text{sign}(Z_\uparrow Z_\downarrow) \sqrt{\frac{T_\uparrow T_\downarrow}{1-\lambda^2}}\sin{\Big(\frac{\sigma\theta +\arccos{\lambda}}{2}\Big)}\,\sin\chi.
\end{align}
Here, $\lambda\equiv \lambda(\chi,0)$ and we introduce the sign convention $\sigma=+(-)$ for spin $\uparrow(\downarrow)$. 
Measurement of the spin-dependent supercurrent, e.g., by coupling the supercurrent to a superconducting microwave 
resonator~\cite{Urbina, Girvin, ASQ3}, can provide a qubit state readout.

\section{ Discussion and conclusion}
By fabricating an array of Josephson junctions, 
our hybrid structure may be generalized to a multi-ASQ system, where the FI magnetizations provide a single-qubit control knob. In a layout with two qubits whose spatial separation is comparable to $\xi$, the
two ABS wave functions with
close energies may hybridize and induce an 
effective exchange interaction between qubits. 
One way to control this interaction could be to consider a single-channel semiconducting
nanowire as the weak link so that a gate voltage might be used
to tune the ABS wave functions (via affecting the transmission of the wires) and therefore manipulate the ABS hybridization~\cite{Nazarov,Nz}. Moreover, in SQUID loops containing these qubits, the interaction between two or multiple qubits can be realized by means of tunable inductive coupling between these SQUID loops~\cite{talk1,*talk2}. This tunable interaction allows single- or two-qubit operations by preventing unwanted qubit crosstalk. 
As an immediate application of a tunable exchange interaction, 
the single-qubit protocol discussed in Eq.~\eqref{steps} can be generalized to a two-qubit geometric gate by periodic manipulation of the exchange interaction~\cite{Xue}. In a multi-ASQ system, 
it would be an interesting future study to
explore other spin-qubit encodings~\cite{Petta,Simon1,*Simon2,*Simon3} that further reduce coupling of the qubit to the environment 
by storing information in the shared spin space of many qubits. 

In this article, we propose a spin qubit in an S-N-FI hybrid structure based on the spin-resolved Andreev-bound states localized in the N region. The qubit 
state can be coherently controlled by manipulating the FI magnetization dynamics. Our hybrid structure demonstrates the potential for 
spintronic methods to control the spin qubits.  We show that a simple protocol for the magnetization dynamics leads to the realization of a nonadiabatic geometric gate.

This work was supported by the U.S. Department of Energy, Office of Basic Energy Sciences under Award No. DE-SC0012190.

\bibliography{Ref}% Produces the bibliography via BibTeX.

\end{document}